\documentclass[a4paper]{jpconf}

\usepackage{graphicx}

\begin{document}

\title{On the possibility of research the photon-photon interaction at the European X-ray Free Electron Laser~-- European XFEL
      }

\author{A N Popov$^{1}$, S V Bobashev$^{1}$,  N O Bezverkhnii$^{1}$, A A Sorokin$^{1,2}$}

\address{$^1$ Ioffe Institute, Saint Petersburg, Russia \\
$^2$ DESY, Hamburg, Germany}


\ead{alexander.popov@mail.ioffe.ru}

\begin{abstract}
The possibility of performing the experimental research in the field of fundamental physics based on the unique instrument~-- European X-ray Free Electron Laser (E-XFEL) is considered in this paper. The calculations of the reaction $\gamma + \gamma \to e^{+} + e^{-}$  cross section for gamma quanta with $E \sim~(1-100)\,\mbox{GeV}$ energy with X-ray photons are performed. The possibility of experimental registration of reaction $\gamma + \gamma \to e^{+} + e^{-}$ product are reviewed. Also,  the optical depth of the interaction between gamma-rays with the E-XFEL's photon pulses is estimated. Astrophysical applications are discussed.
\end{abstract}

\section{Introduction}
The scattering of two photons is a well-known quantum process which exhibits the non-linear nature of quantized electromagnetic fields. Although light-by-light scattering was predicted almost 100 years ago its direct experimentally observation still remains to be discovered.Until today, several ways of directly detected this phenomenon in the laboratory have been proposed. For example, Compton-backscattered photons against laser photons \cite{7}, collisions of photons from microwave waveguides \cite{8} or high-power lasers \cite{9,10,11}, and at photon colliders \cite{12,13,holeraum}. In \cite{LHCPHPH} using computer simulations, the authors show that lead-lead collisions provide the opportunity for seeing a scattering of quasireal photons in the LHC. Direct observation of photon-photon scattering is worth the effort because of any deviation from the predicted could be evidence of new physics, e.g., the existence of supersymmetry. In this paper, we discuss the possibilities of conducting experimental research in the field of fundamental physics based on the E-XFEL.

The E-XFEL is the international research project which involved 12 states. It was officially launched on September 1, 2017.  Compared with modern a synchrotron light sources, the peak brightness of the E-XFEL is more than 8 orders of magnitude higher, the radiation has a high degree of transverse coherence and pulse duration is about 10 fs. The particularities of the E-XFEL's radiation make it possible to conduct groundbreaking research in fundamental physics. Table 1 shows the X-ray beam parameters of one of the three channels of the E-XFEL. 

\begin{center}
\begin{table}[h]
\caption{\label{opt}Radiation parameters of one of the E-XFEL's channels SASE3}
\centering
\begin{tabular}{@{}*{4}{l}}
\br
Parameter&\multicolumn{3}{c}{SASE3}\\
\mr
Energy $e^-$ (GeV) & $17.5$ & $17.5$ & $10.0$\\
Wavelength (nm) & $0.4$ & $1.6$ & $6.4$\\
Eneregy $\gamma$ (keV) & $3.1$ & $0.8$ & $0.2$\\
Peak power (GW)  & $80$ &  $130$ & $135$\\
Average power (W)  & $260$ & $420$ & $580$\\
FWHM ($\mu m$) & $60$ & $70$ & $95$ \\
Beam divergence ($\mu rad$) & $3.4$ & $11.4$ & $27$\\
Coherence time (fs) & $0.34$ & $0.88$ & $1.9$\\
Spectral width (\%) & $0.2$ & $0.3$ & $0.73$\\
Pulse duration (fs) & $100$ & $100$ & $100$\\
Photons in impulse  & $1.6 \cdot 10^{13}$ & $1.0 \cdot 10^{14}$ & $4.3 \cdot 10^{14}$\\
Average flux (phot/s) & $5.2 \cdot 10^{17}$ & $3.4 \cdot 10^{18}$ & $1.4 \cdot 10^{19}$\\
Peak brilliance & $2.0 \cdot 10^{33}$ & $5.0 \cdot 10^{32}$ & $0.6 \cdot 10^{32}$\\
Average brilliance & $6.4 \cdot 10^{24}$ & $1.6 \cdot 10^{24}$ & $2.0 \cdot 10^{24}$\\
\br
\end{tabular}
\end{table}
\end{center}

\begin{figure}[b]
\begin{minipage}{16pc}
\includegraphics[width=14pc, angle=270]{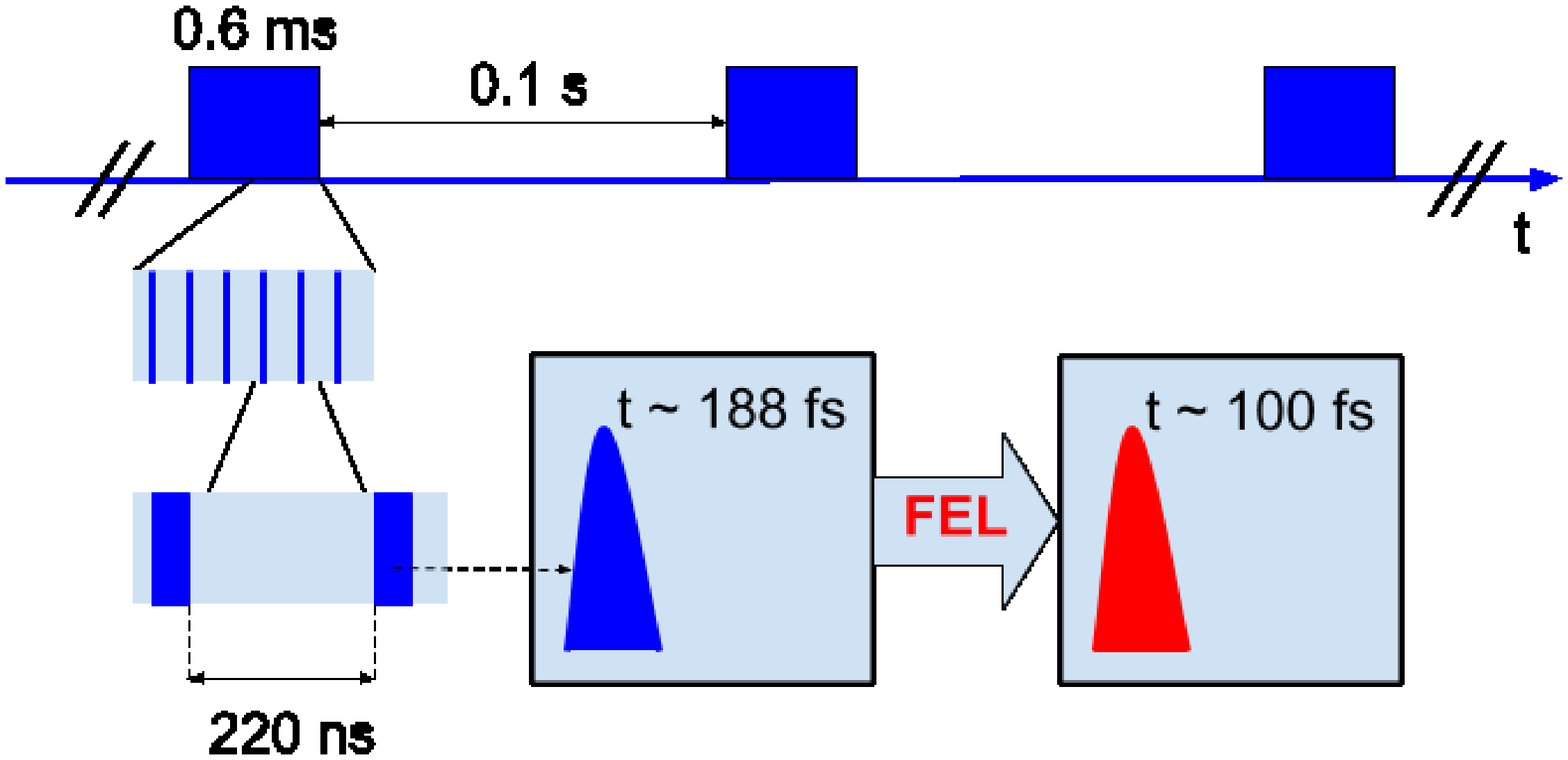}
\caption{\label{fig_structure} 
E-XFEL's radiation time structure.} 
\end{minipage}
\hspace{2pc}%
\begin{minipage}{16pc}
\includegraphics[width=16pc]{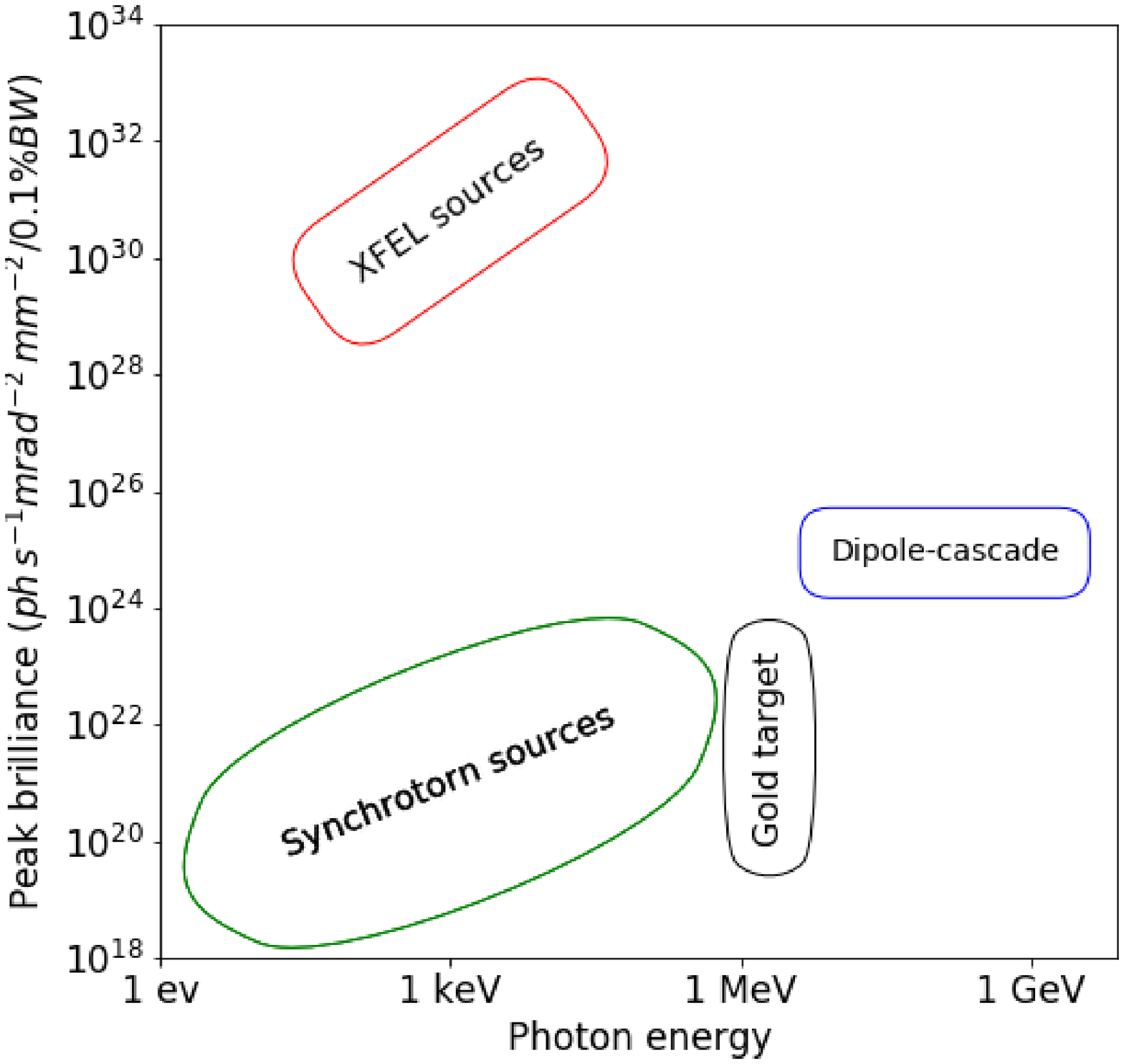}
\caption{\label{fig_structure} 
Comparison of gamma photon sources according to data from \cite{PhysRevX.7}. The "Gold target" is the bremsstrahlung gamma-ray source from \cite{holeraum}.} 
\end{minipage}
\end{figure}

\section{Experimental proposal and possible gamma-ray sources}

The experimental proposal in simplified form is as follows. A high-energy photon beam, generated via the gamma-ray source is fired into a vacuum channel where it intersects with the E-XFEL's x-ray photon beam.  The scheme takes advantage of the high photon densities that are characteristic of the radiation fields produced by E-XFEL. The vacuum channel is surrounded by multi-layer charged particle detectors for detecting elector-positron pairs. The detectors are protected from external radiation and cosmic rays.

One of the central problems of this proposal is the gamma-ray source. In the work \cite{holeraum}, it is proposed to use bremsstrahlung radiation of ultrarelativistic electrons with an energy of about GeV passing through a gold target to create a gamma-ray beam. Electrons and positrons emitted from the back surface of a gold target deviate from the interaction region using a magnetic field.

Another way is proposed in a recent paper \cite{PhysRevX.7}. Electromagnetic dipole cascades can become one of the possible sources of gamma. The paper demonstrated the possibility of creating a stable directional source of photons of GeV energies. Using 3D QED particle-in-cell modeling and analytical evaluations, the authors show that it is possible to attain a peak power of 10 PW. When this becomes possible, there will be a real possibility to carry out this kind of experiments. Fig. 2 shows the approximate location of possible gamma-ray sources on a brilliance-energy map.

\section{The optical depth estimation}
The temporal structure of the E-XFEL laser, shown in Fig. 1, is a sequence of trains consisting of 2700 femtosecond pulses. The total repetition rate of the trains is 10 Hz. For one train, the optical depth becomes equal to $5.4 \cdot 10^{-4}$, that is, when the gamma-quantum crosses 100 trains, the optical depth becomes equal to 1. 

The cross section of interaction of gamma quantum 
to thermal photon with producing electron-positron pair is 
\cite{Gould1967_cross_section}:
\begin{equation}
\sigma = \frac{\pi}{2} \, r_{e}^{2} \, (1 - v^{2}) \,
         \left(
               (3 - v^{4}) \, 
               {\mathrm{ln}}
               \left( \frac{1+v}{1-v} \right)
             - 2\, v \, (2 - v^{2})
         \right)
         \, h(s)
\end{equation}
where $r_{e} = \frac{e^{2}}{m c^{2}}$
is classical electron radius, 
$m$ is mass of electron,
$h(s)$ is Heaviside function
($h(s) = 1$ at $s>0$ and $h(s) = 0$ at $s<0$),
\begin{equation}
v = \sqrt{ 1 - 1/s \, }
\mbox{\ \ and \ \ }
s = \frac{ E \epsilon }{ m^{2} c^{4} }
     ( 1 - \cos\Psi )
,
\label{eqn_1}
\end{equation}
$E$ is energy of gamma quantum,
$\epsilon$ is energy of thermal photon, 
radiated by intracluster gas,
$\Psi$ is angle between its impulses.

Consider a situation where photons move towards each other, i. e. $\theta = \pi$. the X-ray beam in XFEL bunch reaches 100 photons on square $10^{-16} \mbox{ cm}^2$ for impulse in order 10 fs. It remains to make a closer study of number of photons are per area, equal to area of the reaction cross section. Put in (\ref{eqn_1}) the product of the energies $\epsilon E$ equal to $1.2 m_e^2 c^4$. Then (for $\theta = \pi$) $\sigma \approx 2 \cdot 10^{-25} \mbox{ cm}^2$. This means that the optical depth $\tau \approx 2 \cdot 10^{-7}$ is collected for one pulse. The magnitude of the cross section depends on the angle of interaction (Fig. 3). Figure 4 shows the dependence of the reaction cross section on the energy of the oncoming gamma photon.

\begin{figure}[h]
\begin{minipage}{18pc}
\includegraphics[width=16pc]{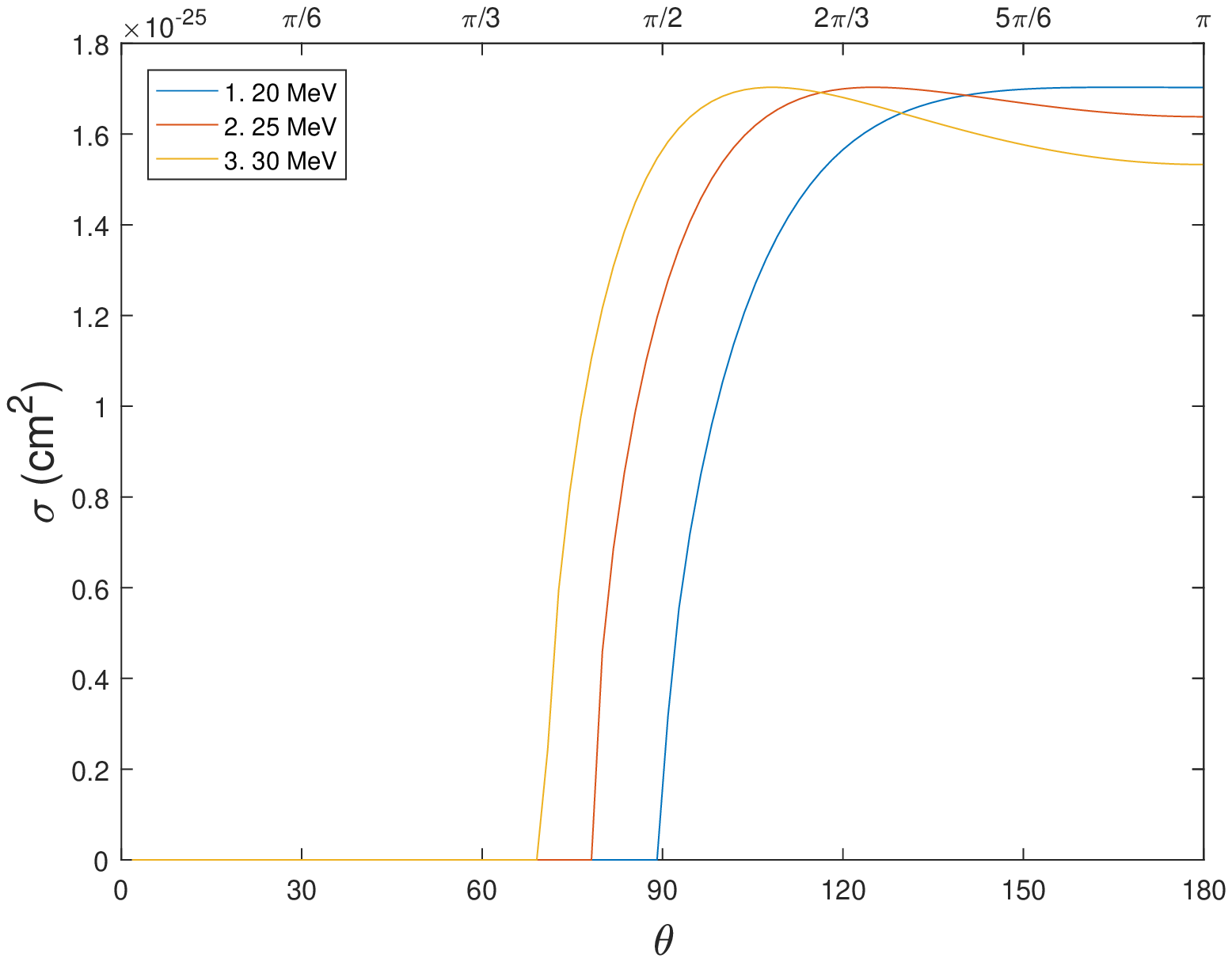}
\caption{\label{fig_v6_L}
The dependence of  cross-section $\sigma$
on the angle between beams for X-ray photon energy $\epsilon = 25 \mbox{ keV}$ and for three values of gamma photon energy.
        }
\end{minipage}
\hspace{2pc}%
\begin{minipage}{18pc}
\includegraphics[width=16pc]{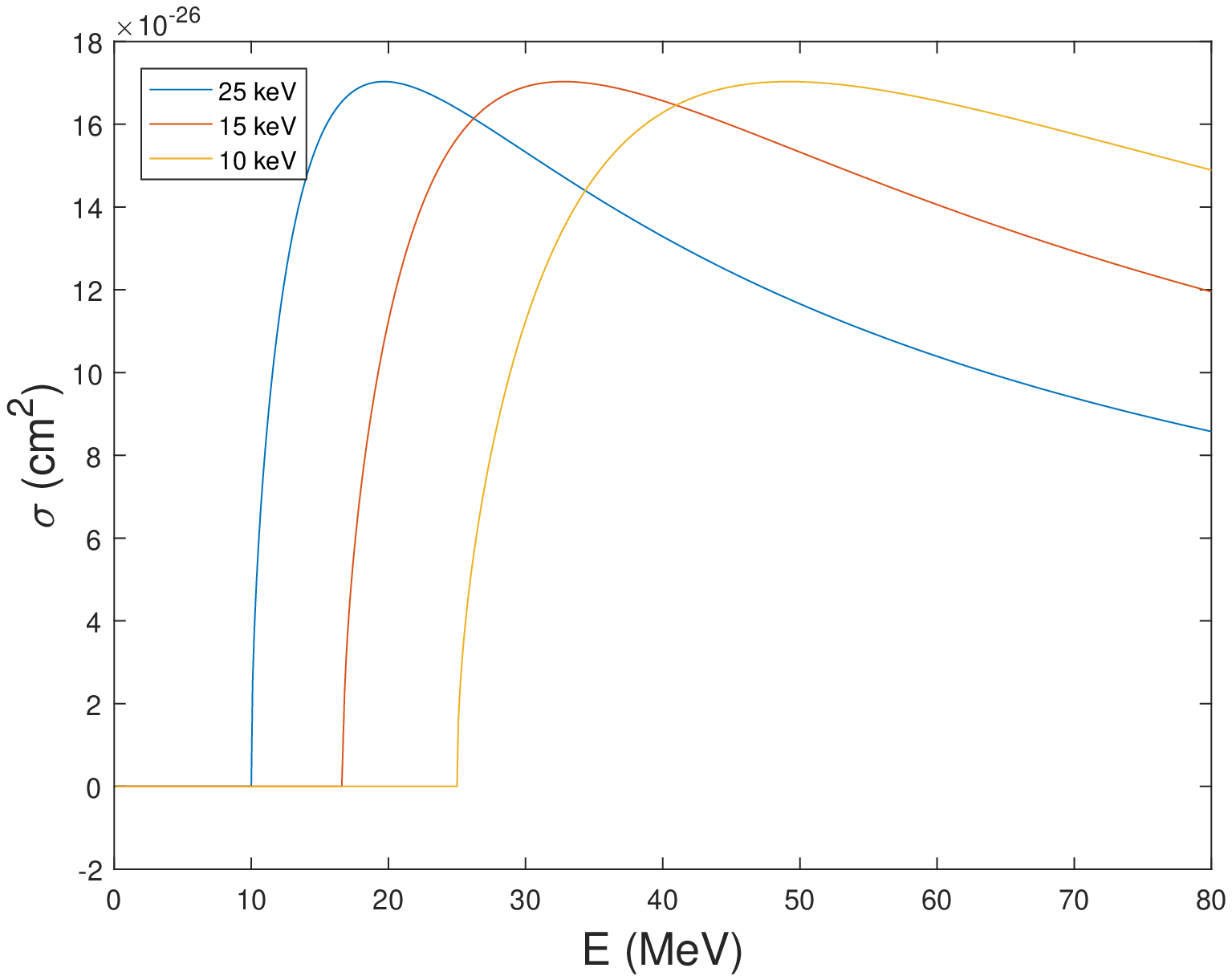} 
\caption{\label{fig_v6_L}
The dependence of cross-section $\sigma$
on gamma photon energy with three values of X-ray photon energy.
        }
\end{minipage}
\end{figure}

%

\section{Astrophysical applications}

Another reason to research this problem are numerous astrophysical applications. For example, two-photon pair-production processes are one of the mechanisms of electron-positron plasma generation in the popular outer gap model of magnetospheric emission from pulsars \cite{MonthlyNotices}. 

Another example is gamma-ray emission from cosmological sources.  The observed gamma-ray spectrum of them contains information about physical conditions and composition of matter in the Universe at various stages of its evolution. 
In this case, it is very important to take into account the factors leading to spectral distortions as absorbtion due to the interaction with background photons. The process of absorption of gamma rays with the production of electron-positron pairs in the collision of them with cosmological background photons~\cite{ruffini} makes the most important contribution to distortion of spectrum in the gamma-ray range. 
 For gamma quanta with energies $E \sim 100 \mbox{ GeV} \div 100 \mbox{ TeV}$, this is mainly interaction with clusters of galaxies, which are the largest gravitationally bound objects in the Universe.

The noticeable part of the baryon component of the cluster is a rarefied ($n \sim 10^{- 3} \div 10^{-2} \mbox{cm}^{-3}$) hot ($T \sim 1 \div 10 \mbox {keV}$) intracluster gas~\cite{ruffini,vikhlinin2006}.
Moreover, due to the bremsstrahlung of electrons of this gas, clusters are also transformed into huge photon reservoirs~\cite{lea}. Therefore, in addition to the components of the intergalactic background, distortions in the spectrum of distant cosmological sources of gamma radiation can also be caused by the interaction of gamma-rays with photons radiated
by gas in cluster of galaxies.

The authors of \cite{popov} have found out that the interaction with the photons of the bremsstrahlung of gas in clusters of galaxies makes a small contribution to the optical depth, compared with absorption by microwave, infrared, visible and radio photons, for almost all gamma-ray energies. However, for energies $E \sim 1-100 \, \mbox{GeV}$, the scattering effect on photons of the bremsstrahlung galaxies can dominate and amounts to $\tau \approx 10^{-5}$.

\section{Conclusions} 

In conclusion, two-photon pair-production processes should easily be observable at this kind experiment on E-XFEL. This experiment can be considered both as a test of the standard model and as a simulation of astrophysical processes. It is a good possibility to research the absorption of gamma radiation in the laboratory.
However, the implementation of the experiment described in this article
is a task with many problems:
\begin{enumerate}
\item The real source of gamma-ray.
\item The photon interaction geometry.
\item Special aspects of the interaction of photons with coherent E-XFEL's radiation.
\end{enumerate}
These problems require further study and will be considered in the further
 papers.

\ack
We sincerely thank D. P. Barsukov for help, comments and useful discussions.

\end{document}